# HAT-P-26b: A Neptune-Mass Exoplanet with a Well Constrained Heavy Element Abundance

Hannah R. Wakeford,[1*†] David K. Sing,[2†] Tiffany Kataria,[3] Drake Deming,[4] Nikolay Nikolov,[2] Eric D. Lopez,[1,5] Pascal Tremblin,[6] David S. Amundsen,[7,8] Nikole K. Lewis,[9] Avi M. Mandell,[1] Jonathan J. Fortney,[10] Heather Knutson,[11] Björn Benneke,[11] and Thomas M. Evans[2]

[1]NASA Goddard Space Flight Center, 8800 Greenbelt Road, Greenbelt, MD 20771, USA.
[2]Astrophysics Group, University of Exeter, Physics Building, Stocker Road, Exeter, Devon, EX4 4QL UK.
[3]NASA Jet Propulsion Laboratory, 4800 Oak Grove Drive, Pasadena, CA 91109, USA.
[4]Department of Astronomy, University of Maryland, College Park, MD 20742, USA.
[5]Institute for Astronomy, Royal Observatory Edinburgh, University of Edinburgh, Blackford Hill, Edinburgh, UK.
[6]Maison de la Simulation, Commissariat à l'énergie atomique et aux énergies alternatives (CEA), CNRS, Université Paris-Sud, Université Versailles Saint-Quentin-en-Yvelines (UVSQ), Université Paris-Saclay, 91191 Gif-sur-Yvette, France.
[7]Department of Applied Physics and Applied Mathematics, Columbia University, New York, NY 10025, USA.
[8]NASA Goddard Institute for Space Studies, 2880 Broadway, New York, NY 10025, USA.
[9]Space Telescope Science Institute, 3700 San Martin Drive, Baltimore, MD 21218, USA.
[10]Department of Astronomy and Astrophysics, University of California, Santa Cruz, CA 95064, USA.
[11]Division of Geological and Planetary Sciences, California Institute of Technology, Pasadena, CA 91125, USA.

*Correspondence to: stellarplanet [at] gmail [dot] com

† These authors contributed equally to this work.

## Abstract

**A correlation between giant-planet mass and atmospheric heavy elemental abundance was first noted in the past century from observations of planets in our own Solar System, and has served as a cornerstone of planet formation theory. Using data from the Hubble and Spitzer Space Telescopes from 0.5 to 5μm, we conducted a detailed atmospheric study of the transiting Neptune-mass exoplanet HAT-P-26b. We detected prominent $H_2O$ absorption bands with a maximum base-to-peak amplitude of 525ppm in the transmission spectrum. Using the water abundance as a proxy for metallicity, we measured HAT-P-26b's atmospheric heavy element content ($4.8^{+21.5}_{-4.0}$ times solar). This likely indicates that HAT-P-26b's atmosphere is primordial and obtained its gaseous envelope late in its disk lifetime, with little contamination from metal-rich planetesimals.**

**HAT-P-26b is a Neptune-mass planet with a lower bulk density as compared with those of the four other Neptune-sized planets with well-measured masses and radii (Uranus, Neptune, GJ 436b, and HAT-P-11b)** (*1*). Neptune-sized worlds are among the most common planets in our galaxy and frequently exist in orbital periods very different from that of our own Solar System ice giants (*2*). Atmospheric studies using transmission spectroscopy can be used to constrain their formation and evolution. The low gravity (4.17 ms$^{-2}$) and moderate equilibrium temperature ($T_{eq} \approx 990$ K) (*1*) of HAT-P-26b results in a large atmospheric scale height, which is ideal for characterization studies that observe the wavelength dependence of the starlight filtered through the atmosphere during a transit.

The atmospheres of Neptune-mass worlds could have arisen from many different sources resulting in a wide range of possible atmospheric compositions. Depending on their formation and evolutionary history, atmospheres rich in H/He, $H_2O$, and $CO_2$ are all expected to be possible (*3*). H/He rich atmospheres are formed if gas accretes directly from the proto-planetary disc. Alternatively, many of these planets could be water-worlds with an $H_2O$-rich atmosphere, or a rocky planet with an atmosphere produced by outgassing. For hot neptunes in particular, it is an open question as to whether these exoplanets contain large amounts of water and other ices, and how much of that is mixed into the detectable atmospheric envelope. Previous observations of Neptune-mass exoplanets show both cloudy atmospheres, such as that of GJ 436b (*4*), and relatively clear atmospheres as seen in HAT-P- 11b (*5*), where a muted $H_2O$ absorption band was detected.

A correlation between giant planet mass and atmospheric elemental abundance was first measured from the $CH_4$ abundance in the atmospheres of Jupiter (*6*), Saturn (*7*), Uranus (*8*), and Neptune (*9*) and has served as a constraint of planet formation theory (*10*). Abundances of key species have now begun to be measured in exoplanets, such as the well constrained $H_2O$ abundance on the two Jupiter-mass planet WASP-43b (*11*). Atmospheric abundance measurements for Neptune and smaller mass exoplanets remain essentially unconstrained, known only within several orders of magnitude, as the detection of $H_2O$ in HAT-P-11b implies metallicities between 1 to 700× solar (*5*). We add an additional point in the mass-metallicity trend from an observational study of the extrasolar planet HAT-P-26b, which has a similar mass to that of Neptune and Uranus (*1*).

**We observed four transits of HAT-P-26b with the Hubble Space Telescope (HST) via two observational programs:** One transit was observed with the HST Space Telescope Imaging Spectrograph (STIS) (*12*) G750L grating (covering 0.5 to 1.0 μm), and one transit with the HST Wide Field Camera 3 (WFC3) (*13*) G102 grism (0.8 to 1.1 μm). We observed a further two transits using HST WFC3 G141 grism (1.1 to 1.7 μm). We also analyzed two archival Spitzer Infrared Array Camera (IRAC) (*14*) 3.6 and 4.5 μm transits. The full transmission spectrum measured for the atmosphere of HAT-P-26b from 0.5 to 5 μm is shown in Figure 1.

We performed a data reduction using the marginalization methods outlined in previous studies to conduct a uniform analysis (*15-18*). Following standard practice, we monitored each transit with observations occurring before-, during-, and after- transit. We first analyzed the band-integrated light curves in order to obtain the broad-band planet-to-star radius ratio — the radius of the planet ($R_p$) over the radius of the star ($R_*$), ($R_p/R_*$) — by summing the flux across the whole spectral wavelength of each visit. Each light curve is corrected by marginalizing over a grid of systematic models appropriate to each instrument and observational mode (*18*).

From the band-integrated analysis, we obtained a fit for the inclination (*i*), and the ratio of the semimajor axis of the planet (*a*) to the stellar radius ($a/R_*$). Due to the limited phase coverage obtained during each HST visit, caused by its low-Earth orbit, we combined these fits with the previously published values (*1,19*) and calculated the weighted mean. From each band-integrated analysis, we also fitted for the center of transit time. We then fixed each of these parameters for each visit in the subsequent spectroscopic light curve analysis. We incorporated the eccentricity and fixed it to a non-zero value (*1*). Each of these parameters is given in table S1, along with the general system properties and derived parameters from this analysis (*18*).

To create the transmission spectrum (Fig. 1 and Table 1), we extracted various wavelength bins for the HST STIS and WFC3 spectra and separately fitted each bin for $R_p/R_*$ and detector systematics (*18*). We measured the WFC3 G141 transmission spectrum separately for each visit before combining them into a single atmospheric spectrum from 1.1 to 1.7 μm. We rescaled the uncertainties for each data point according to the standard deviation of the residuals and measured any systematic errors correlated in time were measured using the binned residuals (*16*, *17*). This resulted in a final atmospheric transmission spectrum separated into seven STIS bins with an average standard deviation of normalized residuals (SDNR) of 426 parts per million (ppm), seven WFC3 G102 bins with an average SDNR of 342 ppm, 11 WFC3 G141 bins (for each visit) with a SDNR of 262 ppm, and two Spitzer photometric light curves with SDNRs of 381 and 406 ppm in the 3.6 and 4.5μm channels, respectively (fig. S1).

**Using HST WFC3 G141 observations of HAT-P-26b, we measured a distinct H$_2$O absorption feature centered at 1.4 μm, with a base-to-peak amplitude of 525±43 ppm at a confidence of 8.8σ, with its relatively low density indicating that the planet has a considerable H/He envelope.** Measurements in the optical with HST STIS and near-infrared WFC3 G102 suggest the presence of an absorbing cloud deck, which matches well with previously published optical data (fig. S2) (*19*).

To determine the atmospheric structure given the observational measurements, we used (*18*) the one-dimensional forward model ATMO (*20*, *21*), which uses the correlated-k method for radiative transfer (*22*), coupled to a CHNO-based chemical network (*23*) solving for the pressure-dependent abundances of 166 species. We ran 12 ATMO model combinations [which we denote M1-M12; (table S2)] including fixed or varying the C/O, isothermal models or fitting the temperature-pressure (T-P) profile, or using a free-chemistry model in which the abundances of CO, CO$_2$, CH$_4$, and H$_2$O were freely fit, each combination with and without cloud opacities (*18*). We used Bayesian model averaging (BMA) with the Bayesian information criterion (BIC) to combine the model results and incorporate the uncertainty in model selection. Our best-fitting models contained equilibrium chemistry and four fit parameters that consist of the atmospheric temperature, metallicity, cloud contributions, and baseline planetary radius (figs. S3 and S4); they differ in whether they have a fixed or varying C/O. A uniform scattering cloud, parameterized by a gray scattering cross section, was found to provide the optimal solution, which maintained the H$_2$O amplitude feature observed in the WFC3 G141 data (Fig. 1). ATMO models without cloud opacities were largely disfavoured by the data and have low statistical weights (table S2). We additionally modeled the transmission spectrum without the STIS data and found that the use of a uniform scattering cloud does not affect the final abundance measurement (*18*). For the allowed fitting range, we imposed a lower limit on the mixing ratios and T-P profile parameters following (*24*) and an upper limit to the metallicity relative to solar of $10^{4.5}$, and imposed a lower limit of the C/O ratio of 0.01; as only $H_2O$ features are present in the spectra, the abundances of carbon-bearing molecules are largely unconstrained (fig S5) (*18*, *25*). The BMA $H_2O$ abundance is found to have a marginalized value of 4.8× solar and ~1σ (68.2%) uncertainty range of 0.8 to 26× solar (table S2).

To estimate any possible cloud absorbers in the atmosphere of HAT-P-26b, we calculated the global temperature structure using the coupled radiation/circulation model SPARC/MITgcm (*18*, *26-28*). Shown in Fig. 2 are the T-P dependent profiles across different regions of the atmosphere and the condensation curves for various cloud species (*29*). This indicates that the cloud base implied by the transmission spectrum is likely composed of sulfur-based species if the clouds are formed through condensation chemistry. We used $Na_2S$ as an example absorber and calculated (*29*, *30*) the extent of a potential condensate cloud in the atmosphere of HAT-P-26b. We define the cloud top as the height at which the optical depth ($\tau$) is equal to 0.1 (*18*), which is within the pressure range constraints from our transmission spectroscopy measurements. We do not consider the radiative effects of the clouds, which likely produces the warmer than predicted atmospheric temperatures.

Similar to previous studies (*5*, *11*), we used the $H_2O$ abundance as a proxy for HAT-P-26b's atmospheric metallicity, because it is the only spectroscopically abundant species and under equilibrium conditions scales approximately linearly with metallicity. We show in Figure 3 the observed trend in mass-metallicity from the giant planets in our Solar System (*6-9*), where the metallicity is defined by the abundance of methane. We include the measured metallicities of three exoplanets — WASP-43b (*11*), WASP-12b (*30*), and HAT-P-11b (*5*) — on the basis of $H_2O$ abundance and show the range for the revised trend when including the published exoplanet results. The combination of broad wavelength coverage with HST and Spitzer, and strong $H_2O$ absorption, results in an atmospheric metallicity for HAT-P-26b, which sits ~1σ below the combined mass-metallicity trend.

**The metallicity derived for HAT-P-26b lies below the trend observed in Solar System giant planets, suggesting different formation and/or evolutionary processes.** Thermal evolution models show that at fixed H/He envelope fractions, planetary radii shows little dependence on mass for planets with more than ~1% of their mass in the envelope (*31*). For HAT-P-26b we found that it would require a H/He envelope of $21\%_{-4}^{+7}$ for a core of 10% rock and 90% water. Given a relatively massive core, it is not expected that more than a few percent of its envelope will have been lost through photo-evaporation (*32*). Population-synthesis models predict a large diversity in planet accretion histories (*10*), depending on where and when planets accreted their envelopes.

Because of the low heavy-element abundance in the atmosphere, we conclude that the gaseous envelope of HAT-P-26b is primordial. The low metallicity suggests that most of the planet's heavy elements are contained in its core and that the planet's gaseous envelope has not been substantially polluted by planetesimals after it accreted, or is at least less polluted than other planets with similar masses (*10*). This is more likely if the planet formed closer to the star, where it is too hot for ices to form and the solid abundance is lower, particularly for carbon and oxygen, and/or accreted its envelope late in the disk lifetime after most of the planetesimals have been cleared out (*10*, *33*). Such a formation scenario is consistent with recent envelope accretion models (*34*), which argue that most hot neptunes accrete their envelopes in situ shortly before their disks dissipate.

**Acknowledgments:** This work is based on observations made with the NASA/European Space Agency (ESA) Hubble Space Telescope, obtained at the Space Telescope Science Institute (STScI), which is operated by the Association of Universities for Research in Astronomy under NASA contract NAS 5-26555. The HST and Spitzer data are available at http://mast.stsci.edu under project numbers GO-14110 (PI:Sing) and GO- 14260 (PI:Deming) and at http://sha.ipac.caltech.edu under project number 90092 (P.I. Jean-Michel Désert). The combined transmission spectrum is available in Table 1. Support for this work was provided by NASA through grants under the HST-GO-14260 program from the STSci. H.R.W. acknowledges support from the NASA Postdoctoral Program, administered by Universities Space Research Association through a contract with NASA. Several authors acknowledge funding from the European Research Council under the European Union Seventh Framework Program: D.K. S., T. K., N. N., and T. E. under grant 336792; E. L. under grant 313014; and P. T. and D. A. under grant 247060-PEPS. The authors thank J. Goyal and B. Drummond for their contributions to the ATMO model framework. The authors thank the anonymous reviewers for their contribution to this paper, including the statistical reviewer who provided an interesting expansion and discussion.

**Supplementary Materials:**

Materials and Methods

Figures S1-S5

Tables S1-S2

References (*35-52*)

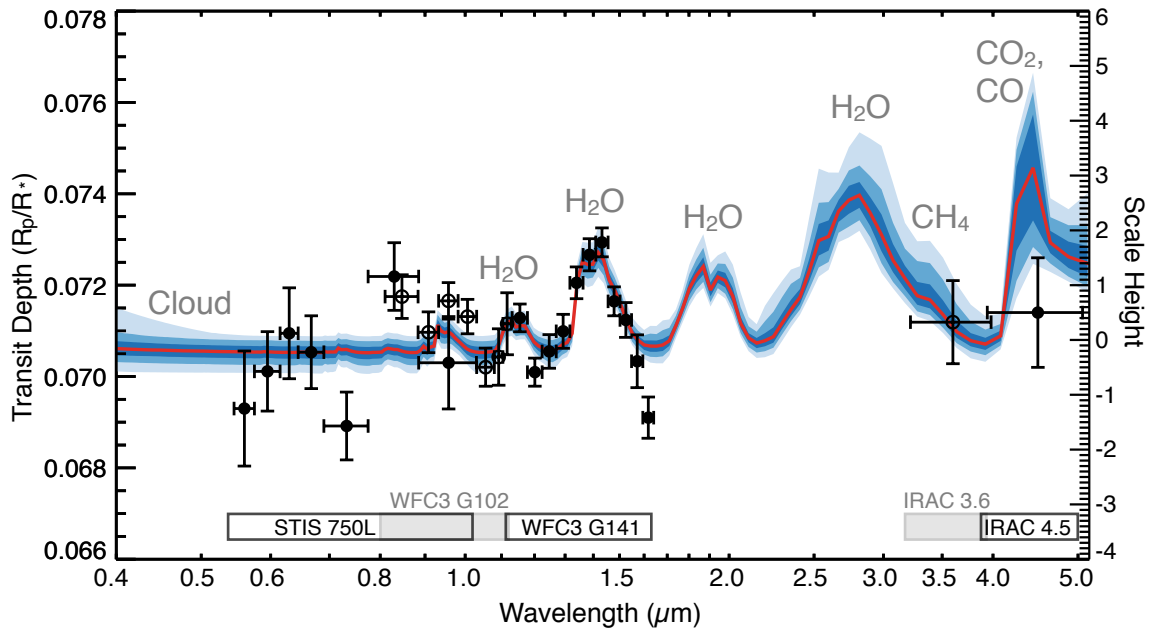

**Fig. 1. The measured transmission spectrum of HAT-P-26b.** We show the atmospheric transmission spectrum (open and solid circles alternating between different observational modes indicated by the labeled bars at the bottom) fitted with a model (red) derived by using the ATMO retrieval code (*18*). The best fitting models have isothermal profiles and include a uniform cloud opacity. Shown here are the results for model M1 with $1\sigma$, $2\sigma$, and $3\sigma$ uncertainty shown in the dark to light blue shaded regions. The right-hand axis shows the corresponding scale of the atmospheric transmission in terms of planetary scale height, which is a logarithmic parameter of the atmosphere based on the planet's temperature, gravity, and mean molecular weight.

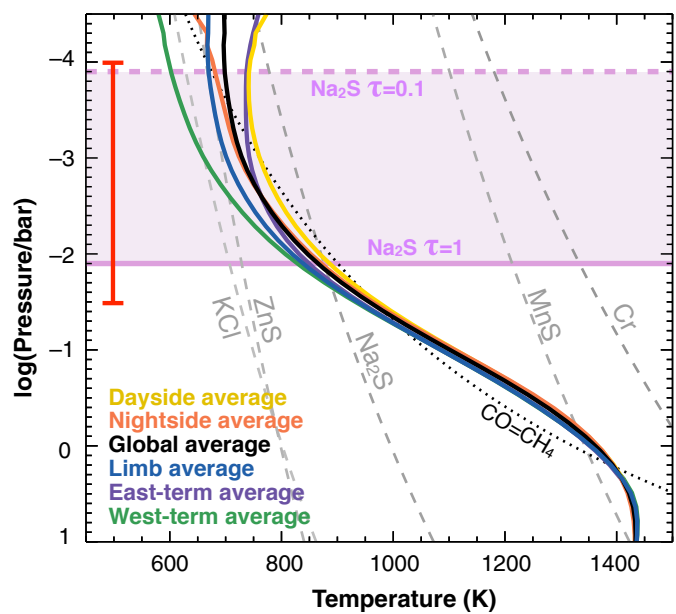

**Fig. 2. Model T-P profiles for HAT-P-26b.** We show the T-P profile for 1×solar metallicity, including global-, dayside-, nightside-, limb-, east terminator-, and west terminator-averaged regions of the atmosphere, as produced from the SPARC/MITgcm code (*27*). Condensation curves for potential cloud species are plotted in gray dashed lines (*29*), where the most likely cloud-forming condensate species are sulfur-based molecules. The red bar shows the pressure range probed with transmission spectral measurements. Using $Na_2S$ as an example, the solid horizontal line indicates the pressure at the base of a $Na_2S$ cloud, and the dashed horizontal line denotes the pressure of the layer where τ becomes 0.1. The CO to $CH_4$ gas transition is shown as a dotted line to indicate where the abundance of CO is equal to $CH_4$.

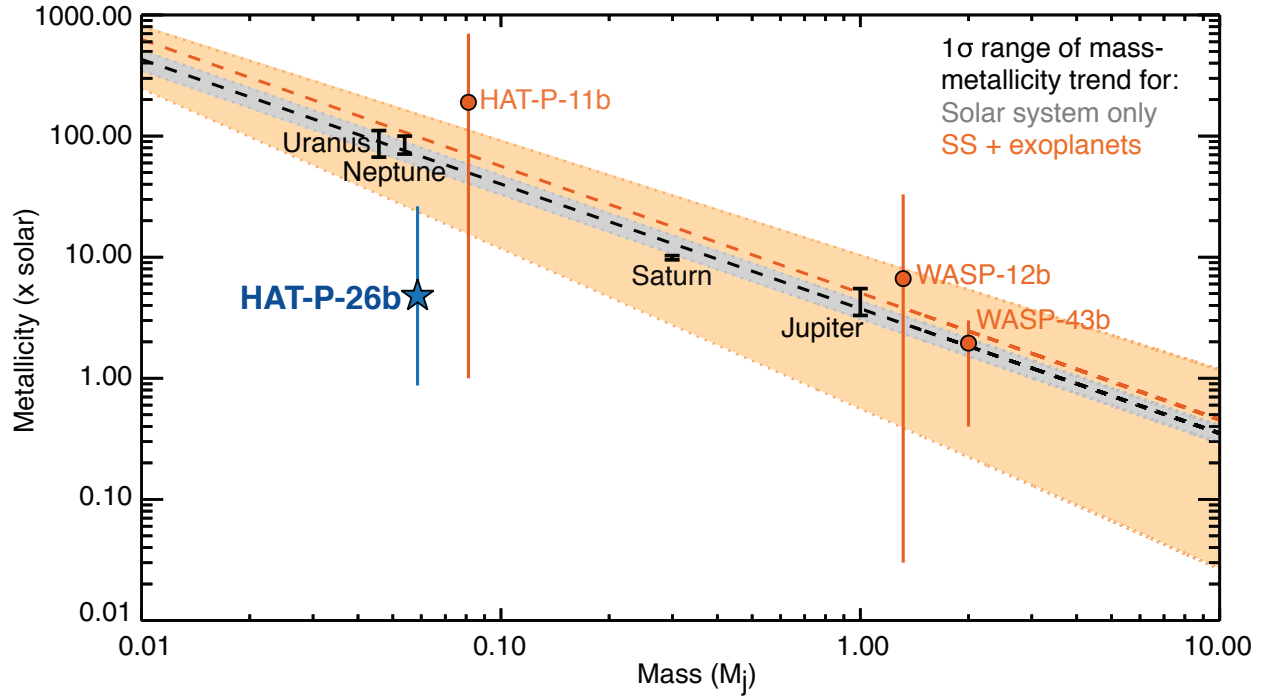

**Fig. 3. The correlation between giant planet mass and atmospheric metallicity.** Mass is in Jupiter masses ($M_J$), and metallicity is in solar units (× solar). Shown are the solar system planets (gray) (*6-9*) and the three previously measured exoplanets WASP-43b, WASP-12b, and HAT-P-11b (orange circles) (*11, 26, 5*), with 1σ error bars. We added the new metallicity measurement for the exoplanet HAT-P-26b (blue star) at $4.8^{+21.5}_{-4.0}×$ solar, derived by using the BMA across all 12 ATMO models. The derived metallicity is likely less than that of Neptune and Uranus and has a median metallicity closer to the gas giants Jupiter and Saturn. Overlain are linear fits to the mass-metallicity relation for the Solar System planets (gray shaded region) and the Solar System plus exoplanet results (orange shaded region).

**Table 1. Transmission spectrum measured for HAT-P-26b.** $R_p/R_*$ is the measured radius ratio between the planet and star during transit for each instrument and mode.

| Wavelength, λ (μm) | Δλ (μm) | $R_p/R_*$ | Uncertainty |
|---|---|---|---|
| HST STIS G750L | | | |
| 0.560 | 0.030 | 0.06929 | 0.00125 |
| 0.595 | 0.040 | 0.07011 | 0.00086 |
| 0.630 | 0.030 | 0.07094 | 0.00099 |
| 0.667 | 0.044 | 0.07053 | 0.00079 |
| 0.732 | 0.084 | 0.06891 | 0.00074 |
| 0.830 | 0.110 | 0.07218 | 0.00074 |
| 0.957 | 0.144 | 0.07030 | 0.00101 |
| HST WFC3 G102 | | | |
| 0.847 | 0.072 | 0.07174 | 0.00047 |
| 0.908 | 0.048 | 0.07096 | 0.00044 |
| 0.957 | 0.048 | 0.07165 | 0.00039 |
| 1.006 | 0.048 | 0.07131 | 0.00037 |
| 1.055 | 0.048 | 0.07020 | 0.00041 |
| 1.092 | 0.024 | 0.07042 | 0.00061 |
| 1.116 | 0.024 | 0.07115 | 0.00067 |
| HST WFC3 G141 | | | |
| 1.154 | 0.046 | 0.07128 | 0.00030 |
| 1.200 | 0.046 | 0.07009 | 0.00030 |
| 1.247 | 0.046 | 0.07054 | 0.00036 |
| 1.293 | 0.046 | 0.07098 | 0.00037 |
| 1.339 | 0.046 | 0.07205 | 0.00034 |
| 1.386 | 0.046 | 0.07266 | 0.00035 |
| 1.432 | 0.046 | 0.07293 | 0.00031 |
| 1.478 | 0.046 | 0.07164 | 0.00031 |
| 1.524 | 0.046 | 0.07123 | 0.00038 |
| 1.571 | 0.046 | 0.07033 | 0.00057 |
| 1.617 | 0.046 | 0.06909 | 0.00045 |

| Spitzer IRAC 3.6 & 4.5 | | | |
|---|---|---|---|
| 3.600 | 0.76 | 0.07119 | 0.00091 |
| 4.500 | 1.12 | 0.07140 | 0.00120 |

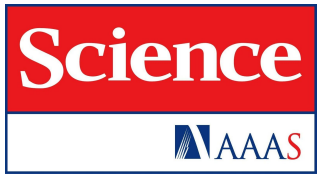

# Supplementary Materials for

HAT-P-26b: A Neptune-mass Exoplanet with a well constrained Heavy Element Abundance

Hannah R. Wakeford,[1*†] David K. Sing,[2†] Tiffany Kataria,[2,3] Drake Deming,[4] Nikolay Nikolov,[2] Eric Lopez,[1,5] Pascal Tremblin,[6] David S. Amundsen,[7,8] Nikole Lewis,[9] Avi Mandell,[1] Jonathan J. Fortney,[10] Heather Knutson,[11] Björn Benneke,[11] and Tom Evans[2]
* Correspondence to: hannah.wakeford@nasa.gov
† These authors contributed equally to this work.

**This PDF file includes:**

Materials and Methods
Figs. S1 to S5
Table S1 to S2

**Materials and Methods**

We observed the transit of HAT-P-26b with HST STIS, and HST WFC3 over four new transit events, and analysed archival data from Spitzer IRAC 3.6 & 4.5 micron for two other transit events. HAT-P-26 is a quiet star with a consistent stellar activity level. This is shown by a consistent transit depth measured across all six observations with no indication of starspot occultations (Figure S1). If there are any unocculted spots the stellar spectrum is unlikely to mimic the observed water feature for starspots of any plausible temperature contrast (*35*). For our HST analysis we follow standard practice and discard the first orbit of each visit as it contains vastly different systematics to the subsequent orbits due to thermal settling of HST after initial target acquisition (e.g. *15-17, 35*).

All the transit light curves were fitted with analytical transit models (*36*). We apply cosmic ray removal using IDL routines from (*15*), and account for stellar limb darkening using a 4-parameter limb-darkening law (*37*). We use marginalization across a series of stochastic models (*38*) to account for differing observatory and instrument configuration systematics. Each light curve is fit using a Lavenberg-Marquart (L-M) least-squares algorithm (*39*). The uncertainties on each data point were initially set to pipeline values dominated by photon and readout noise. After an initial fit the lightcurve, uncertainties are rescaled based on the standard deviation of the residuals, taking into account any underestimated errors calculated by the L-M reduction, which is then re-run to calculate the final fit to the data (*16, 17*). The L-M fit assumes the probability space around the best-fitting solutions are well described by a multivariate Gaussian distribution, as with previous studies (*17, 40-42*) we check this using a Markov Chain Monte Carlo (MCMC) analysis (*43*) and find excellent agreement with our data showing it is normally distributed. The transit depth uncertainty for each systematic model is then determined using the covariance matrix from the second L-M reduction. For each of the lightcurve fits we assume a normal likelihood function. We approximate the evidence-based weight based on the Akaike Information Criterion (AIC) and marginalize across all models to compute the desired light curve parameters and uncertainties (e.g. *16, 17, 40*). Using marginalization across a grid of stochastic models allows us to therefore account for all tested combinations of systematics and obtain robust center of transit times from the band-integrated lightcurve, and transit depths for each spectroscopic lightcurve.

STIS observations and data analysis

We observed one transit of HAT-P-26b on 2016 January 25 using HST STIS. Our STIS observations were conducted using the G750L grating with a wide 52x2" slit to avoid slit losses. The STIS dataset was pipeline-reduced with the latest version of CALSTIS (*12*) and corrected for detector fringing using contemporaneous fringe flats. The spectral aperture extraction was done with the software package IRAF, using a 13-pixel-wide aperture with no background subtraction, which minimizes the out-of-transit standard deviation of the band-integrated light curves. The extracted spectra were then Doppler-corrected to a common rest frame through cross-correlation, which helped remove sub-pixel wavelength shifts in the dispersion direction (*16, 41*). To correct for telescope and instrument systematics we use a grid of stochastic polynomial models accounting for a linear change in flux over the course of the whole observation ($\theta$), orbit to orbit flux corrections in HST orbital period ($\phi$) caused by thermal variations in the telescope, and linearly for the *x* and *y* detector position determined from linear spectral traces in IRAF. We use linear combinations of each parameter, $\theta$, *x*, and *y*, and with $\phi$ up to a 4th order polynomial. This produces a grid of 80 stochastic models for the HST STIS systematics applied to each light curve which are then marginalized over to determine the transit parameters.

We divide the STIS wavelength range into multiple custom bins for different wavelength ranges and measure the $R_p/R_*$ from each spectroscopic light curve following the same procedure as detailed for the band-integrated light curve. Figure S1a shows the spectroscopic light curves for seven custom bins offset for clarity and corrected by the highest weight systematic model, which contained corrections in $\theta$, $\phi^3$, and *y*. We test a series of different bin widths and positions, however, due to the large scatter and relatively low resolution of the STIS observations we use broader bins in the red end of the wavelength regime to avoid increasing the uncertainty. The final scatter observed in the STIS transmission spectrum between 0.7-1.0 μm matches well with previously published optical data (*19*), which measures the spectrum at a higher resolution. Remaining outliers are likely the result of fringing which has been seen in various other datasets (e.g. *15, 16, 41*). We show an expanded view of the optical portion of the HAT-P-26b transmission spectrum in Figure S2a, with the HST data, plotted with the previously observed ground-based results (*19*) for visual comparison.

WFC3 observations and data analysis

We observed HAT-P-26 over one visit with WFC3 G102 on 2016 August 16, and two visits with WFC3 G141, the first on 2016 March 12, the second on 2016 May 2. Observations for both spectroscopic grisms were conducted in forward spatial scan mode. Spatial scanning involves exposing the telescope during a forward slew in the cross-dispersion direction and resetting the telescope position to the top of the scan prior to conducting subsequent exposures. Scans with G102 were conducted at a scan rate of ~0.27 pixels per second with a final scan covering ~28 pixels in the cross-dispersion direction. For both G141 visits we used a scan rate of ~0.55 pixels per second with a final spatial scan covering ~62 pixels in the cross-dispersion direction on the detector.

We use the *IMA* (intermediate IR exposure) output files from the CalWF3 pipeline (*13*) which are calibrated using flat fields and bias subtraction. We extract the spectrum from each exposure by taking the difference between successive non-destructive reads. A top-hat filter is then applied around the target spectrum and all external pixels are set to zero which helps to remove cosmic rays (*42*). The image is then reconstructed by adding the individual reads back together. We extract the stellar spectrum from each exposure with an aperture of ±14 pixels for G102 and ±31 pixels for G141 around a centering profile which was found to be consistent across the spectrum for each exposure for all three observations. For our WFC3 data analysis, we use a grid of stochastic models accounting for $\theta$, $\phi$, and shifts in the wavelength position of the spectrum caused by telescope pointing ($\delta_\lambda$). We use combinations with and without $\theta$, and with or without $\phi$ and $\delta_\lambda$ up to the 4th order. This results in a grid of 50 systematic models which we test against our data (see *17* for a table of systematic models).

We divide the WFC3 wavelength range for each grism into a series of bins and measure the $R_p/R_*$ from each spectroscopic light curve following the same procedure as detailed for the band-integrated light curve. For each visit we test a range of bins widths and wavelength ranges for each visit and determine that the shape of the transmission spectrum is robust. Figure S1b shows each spectroscopic light curve for the G102 visit split into 7 bins ($\Delta\lambda \approx 0.048$ μm) between 0.8-1.1 μm and corrected using the highest weight systematic model, which corrects for $\theta$, $\phi^3$ and $\delta_\lambda^2$. We also show each spectroscopic light curve from both WFC3 G141 visits for 11 bins each ($\Delta\lambda \approx 0.046$ μm) between 1.13-1.67 μm, again corrected by the highest weight systematic model as defined by the data which includes systematic corrections for $\theta$, $\phi^4$ and $\delta_\lambda$. We show the transmission spectrum measured from both visits with WFC3 G141 in Figure S2b with the combined transmission spectrum used for the final atmospheric interpretation.

Spitzer observations and data analysis

The Spitzer IRAC observations were conducted in subarray mode (32x32 pixels). Photometry was extracted from the basic calibrated data cubes, produced by the IRAC pipeline after dark subtraction, flat-fielding, linearization and flux calibration (see *15* and references therein). We performed outlier filtering for hot (energetic) or cold (low-count values) pixels in the data by examining the time series of each pixel and subtracted the background flux from each image. We measured the position of the star on the detector in each image incorporating the flux-weighted centroiding method using the background subtracted pixels from each image, for a circular region with a radius of 3 pixels centered on the approximate position of the star. We extracted photometric measurements from our data using both aperture photometry from a grid of apertures ranging from 1.5 to 3.5 pixels (in increments of 0.1) and time-variable aperture photometry, which resulted in the lowest white and random red noise correlated with the data points co-added in time for both channels. The best result was selected by measuring the flux scatter of the out-of-transit portion of the light curves for both channels after filtering the data for 5σ outliers with a width of 20 data points (*16*). To correct for systematic effects, we used a parametric models incorporating systematics associated with the *x* and *y* positions of the stellar centroid on the detector and linear trend in time (see *15* for details). We then use marginalization across all systematic models to calculate the resultant photometric transit depth for each channel. The photometric light curves for the 3.6 and 4.5 μm channels are shown in Fig. S1c binned in time by ~2 minutes. We again do not find a significant discrepancy between the results from this analysis in comparison to the Spitzer results presented in (*19*) with agreement at the 1.4σ and 1.8σ level for the 3.6 and 4.5 μm channels respectively.

Atmospheric retrieval model description

We use ATMO (*20, 21*) to compute the atmospheric metallicity, temperature and absorption species from the observations. ATMO is a code that computes the one dimensional (1D) T-P profile of an atmosphere in hydrostatic and radiative-convective equilibrium, and can also be used as a post-processing tool to compute the emission and transmission spectra from a given 1D atmospheric profile. The radiative transfer equation with scattering and irradiation is solved iteratively in its integral form, with correlated-*k* coefficients or line-by-line opacities, and the scheme has been benchmarked against the Met Office SOCRATES code (*22*). ATMO includes equilibrium chemistry with condensed species for a given elemental solar or non-solar composition (solved by minimization of the total Gibbs free energy), and out-of-equilibrium chemistry with both mixing and photo-chemistry (*21*). For the chemistry, we include 166 species with all gas phase species included in the C, H, N, O network (*23*) plus ~ 40 other gas/condensed phase species that are relevant for exoplanet and brown dwarf atmospheres. The code has been successfully applied to the study of Y, T, and L brown dwarfs and the atmosphere of exoplanets observed by direct imaging (*20, 21*).

The forward ATMO model was coupled to a Levenberg-Marqardt least squares minimizer (*39*) to initially find an optimal model fit, and a Differential Evolution MCMC analysis (*43*) was then used to measure the posterior distribution (Fig. S3 and Fig. S4) and determine the fit confidence intervals. We ran MCMCs with 10 to 12 chains each with 25,000 to 35,000 steps, and generally we found about 12,000 steps were needed for convergence which was monitored with the Gelman-Rubin statistic. The fit used k-coefficients with 500 bands, and confirmed with 5,000 bands uniformly spaced in wavenumber between 1 cm$^{-1}$ and 50,000 cm$^{-1}$ for the best fit model. For the opacity sources, we include $H_2$, He, $H_2O$, CO, $CO_2$, $CH_4$ and $NH_3$ (see references in *20, 22*). We also considered Na and K as potential sources of opacity, however the lack of evidence for their presence in the data (also see *19*) only provided upper limits to their abundances.

To fit the data we consider a grid of 12 different atmospheric retrievals (labeled M1 to M12) using the ATMO model. For each model fit we assume a normal likelihood function on each datapoint. Each model requires the planetary radius to be a free parameter, in addition we fix or fit for the following parameters. The contribution of each to the number of free parameters is shown in parentheses.

  M1: C/O fixed, isothermal, uniform scattering cloud (+1), metallicity (+1)
  M2: C/O fitted (+1), isothermal, uniform scattering cloud (+1), metallicity (+1)
  M3: free-chemistry (+4), isothermal, uniform scattering cloud (+1)
  M4: C/O fixed, TP model (+3), uniform scattering cloud (+1), metallicity (+1)
  M5: C/O fitted (+1), TP model (+3), uniform scattering cloud (+1), metallicity (+1)
  M6: free-chemistry (+4), TP model (+3), uniform scattering cloud (+1)

M7-M12 then consider each of these models without the cloud opacity included which removes one free parameter from each model listed above. In the free-chemistry models (M3, M6, M9, and M12) the abundances of CO, $CO_2$, $CH_4$, and $H_2O$ were freely fit. The C/O ratio for the free-chemistry model fit is then estimated from the four fit molecules (CO, $CO_2$, $CH_4$, $H_2O$) and does not take into account any other species. We apply a uniform scattering cloud which assumes the cloud is uniform throughout the atmosphere and fit for the opacity of the cloud such that it becomes optically thick at the altitude defined by the measured transmission spectra.

To determine the metallicity from using the information from all retrieval models applied to the data we use Bayesian Model Averaging (BMA) to combine the posterior distributions over all the reasonable models weighted by their evidence. We use the Bayesian Information Criterion (BIC) to approximate the evidence of fit $E_q$ for each model $S_q$ given by the probability of the data $D$ given the model $q$ and is often referred to as the marginal likelihood (*38*), such as,

$$\ln E_q = \ln P(D|S_q) \approx -\tfrac{1}{2} BIC. \quad (S1)$$

Each evidence value is then transformed into a weighting such that each retrieval is assigned a percentage of the overall probability. The weight $Wq$ is calculated as

$$W_q = E_q / \sum_{q=0}^{N_q} E_q, \quad (S2)$$

where $N_q$ is the number of retrieval models fit. The weighting for each model is then used to calculate the weighted mean of all parameters of interest $\alpha_q$ and their associated uncertainty $\sigma\alpha_q$ are used to calculate the marginalized parameter $\alpha_m$ and weighted uncertainty $\sigma(\alpha)$ (*17*):

$$\alpha_m = \sum_{q=0}^{N_q} (W_q \times \alpha_q), \quad (S3), \text{ and}$$

$$\sigma(\alpha) = \sqrt{\sum_{q=0}^{N_q} (W_q [(\alpha_q - \alpha_m)^2 + \sigma^2_{\alpha_q}])}. \quad (S4)$$

We show the derived metallicity and uncertainty for each retrieval in Table S2 with the number of free parameters, the calculated BIC and $W_q$. Using BMA we determine that HAT-P-26b has an atmospheric metallicity of 4.8× solar and 1σ limits of 0.8 to 26× solar.

Atmospheric retrieval model C/O constraints

In the models where carbon is an unconstrained free parameter, we find the retrieval prefers solutions in which the C/O and abundances of non-$H_2O$ species are very low. This has a direct effect on the atmospheric scale height, as it can substantially reduce the atmospheric mean molecular weight compared to solar-abundance scaled compositions. With comparatively lower molecular weights, the transmission spectra retrievals can accommodate lower temperatures near 400 to 600 K (Fig. S5), which is substantially lower than expected from radiative-hydrodynamic equilibrium (Fig. 2). Like clouds, low temperatures can reduce the amplitude of the water feature but do so by directly reducing the atmospheric scale height. At these low temperatures near 600K, we would expect substantial $CH_4$ signatures in the transmission spectra, but none are found. In addition, the mean molecular weight of a 30× solar composition model is about 3 atomic mass units (amu), but with very low C/O atmospheres, the $H_2O$ abundance has to increase up to about 100× solar to produce a mean of 3 amu. Thus, additionally fitting for the C/O gives extremely low metallicities (0.001× solar) and C/O ratios (C/O < 0.0001) which are both about three orders of magnitude below the solar value and has the effect of removing practically all molecular species in the atmosphere apart from those containing H, He and O. The $H_2O$ abundance is then greatly increased extending the tails of the posterior distribution for the models fitting for C/O. In Figure S5 we illustrate from M2 the effects on the retrieved temperature and $H_2O$ abundance posterior distributions when placing a lower-limit prior constraint on the C/O. We find a lower C/O of 0.01 provides a reasonable range to allow the ratio to vary, compared to the solar value of 0.56, and restricts the impact on the temperature. Previous exoplanet $H_2O$ abundance results for WASP-43b and HAT-P-11b (e.g., *5,11*) have fixed the C/O at solar, which we have also assumed in model cases M1, M4, M7 and M10, as no lower-limit constraints could be placed on carbon-bearing molecules.

Atmospheric retrieval model results

We summarize our ATMO retrieval fits in Table S2. Overall we find that models that do not incorporate clouds (M7-M12) are disfavoured and have low overall weights (<2%) in our BMA results. In addition, the cloudy models which freely fit for the chemistry or TP profile give similar $H_2O$ abundances, but generally have lower weights (~1%) as the BIC penalizes the models due to their increased number of free parameters. The freely fit TP profiles (M4-6, M10-12) were consistent with the isothermal approximation. Our best fitting models have isothermal profiles, scattering clouds, and a freely fit radius and metallicity, with C/O which is fixed to solar (M1 - 42.5%) or fit using a prior of C/O > 0.01 (M2 - 44.9%).

While the carbon-bearing molecules are not fully constrained by our data, the upper limits on the abundance of $CO_2$ gives a further indication that the atmosphere does not have a high metallicity. $CO_2$ is well known to be highly sensitive to the atmospheric metallicity (e.g., *25, 44*). Over the wavelength range of our data, $CO_2$ has by far its strongest signature in the Spitzer 4.5 micron bandpass (Figure 1). CO also has a strong spectral signature at those wavelengths; because the two are unresolved in the spectra, the $CO_2$ abundance is highly degenerate with CO. However, the measurements can still rule out strong $CO_2$ signatures as would be expected at high metallicities near that of Neptune. We find $CO_2$ would be expected from equilibrium chemistry to be abundant at volume mixing ratio (VMR) of $2 \times 10^{-3}$, if the overall metallicity were 100× solar. However, our posterior distribution of the best-fitting free chemistry model (M3) limits $CO_2$ to several orders of magnitude below that value with a 68.2% upper bound limit on the $CO_2$ VMR of $5 \times 10^{-6}$, which is consistent with compositions near solar (Figure S4).

We additionally test that our results are independent of the STIS data, and the possibility of a wavelength dependence of the cloud deck, by fitting the transmission spectrum without the STIS measurements (i.e. fitting 0.8-5.0μm, WFC3+Spitzer). To address the wavelength dependence of the cloud opacity we assume Rayleigh scattering. Using the highest weight model, we retrieve a water abundance which is consistent with fitting the whole transmission spectrum. The red-most point of the WFC3 G141 measurements is better fit and is within 1σ of the model. This suggests that any cloud likely becomes transparent at these wavelengths, as is expected (e.g. *45, 46*). However, as the measured water abundance is consistent with fitting the whole transmission spectrum with a uniform cloud, described by a single parameter, a more complex wavelength dependent cloud model is not justified in this case.

In summary, using the BIC as an approximation of the evidence we show that M1 and M2 fit the data equally well and are the best fitting models. Using BMA we incorporate the derived metallicities of all tested models to produce a marginalised metallicity value and uncertainty coherent across all models. We additionally test this result using the SCARLET retrieval code (*5*) and find a 1σ bound of 1-25× solar, which agrees well with the ATMO results and final marginalised value. The wide wavelength coverage we obtain with HST and Spitzer, compared with the single $H_2O$ absorption feature detection using WFC3 G141 alone, acts to further constrain the clouds and place strong constraints on the metallicity by the models.

Atmospheric profile and cloud models
The best fitting atmospheric models for the measured transmission spectrum suggest the presence of an absorbing cloud deck at optical wavelengths, which we expect to be produced by a cloud composed of relatively large particles which scatter the light uniformly at multiple wavelengths, thereby hiding any atomic features. To estimate the possible cloud absorbers in the atmosphere of HAT-P-26b, we also calculate the three-dimensional (3D) temperature structure using the SPARC/MITgcm. The SPARC/MITgcm couples the MITgcm, a finite-volume code that solves the 3D primitive equations on a staggered Arakawa C grid (*47*) with a radiative transfer code SPARC that is a two-stream adaptation of a multi-stream radiative transfer code for solar system planets (*48*). The radiative transfer code employs the correlated-k method with 11 bands optimized for accuracy and computational efficiency. The opacities are calculated assuming local thermodynamic and chemical equilibrium (*44*). This code has been used extensively to model the atmospheric circulation of hot Jupiters, hot Neptunes and super Earths (e.g., *26-28, 49*). We use 1× solar abundance values and utilize the system parameters listed in Table 1. We compare the T-P profiles to condensation curves (*49*) for various species between 500 and 1500 K to determine the most likely condensation cloud species responsible for the observed transmission spectrum (Figure 2).

To calculate the vertical extent of potential $Na_2S$ clouds on HAT-P-26b, we utilize the cloud code of (*50*), updated for $Na_2S$ cloud formation (*29*). We denote the cloud base as the region where the limb-averaged TP profile intersects the condensation curve of $Na_2S$ (Figure 2) and the cloud top as the region of the limb where the slant optical depth equals one (*51*). Previous studies have shown that $Na_2S$ is highly scattering (*29, 46*) with a high enough abundance to form optically thick clouds in exoplanet atmospheres (*29*). The pressure extent of $Na_2S$ condensate derived from the cloud model is in agreement with the pressure range probed by transmission spectral observations. Additionally, as the most likely cloud to condense at this pressure is $Na_2S$, it is likely that all of the Na is locked up the condensed phase and would not present as atomic lines produced by sodium in the gas phase.

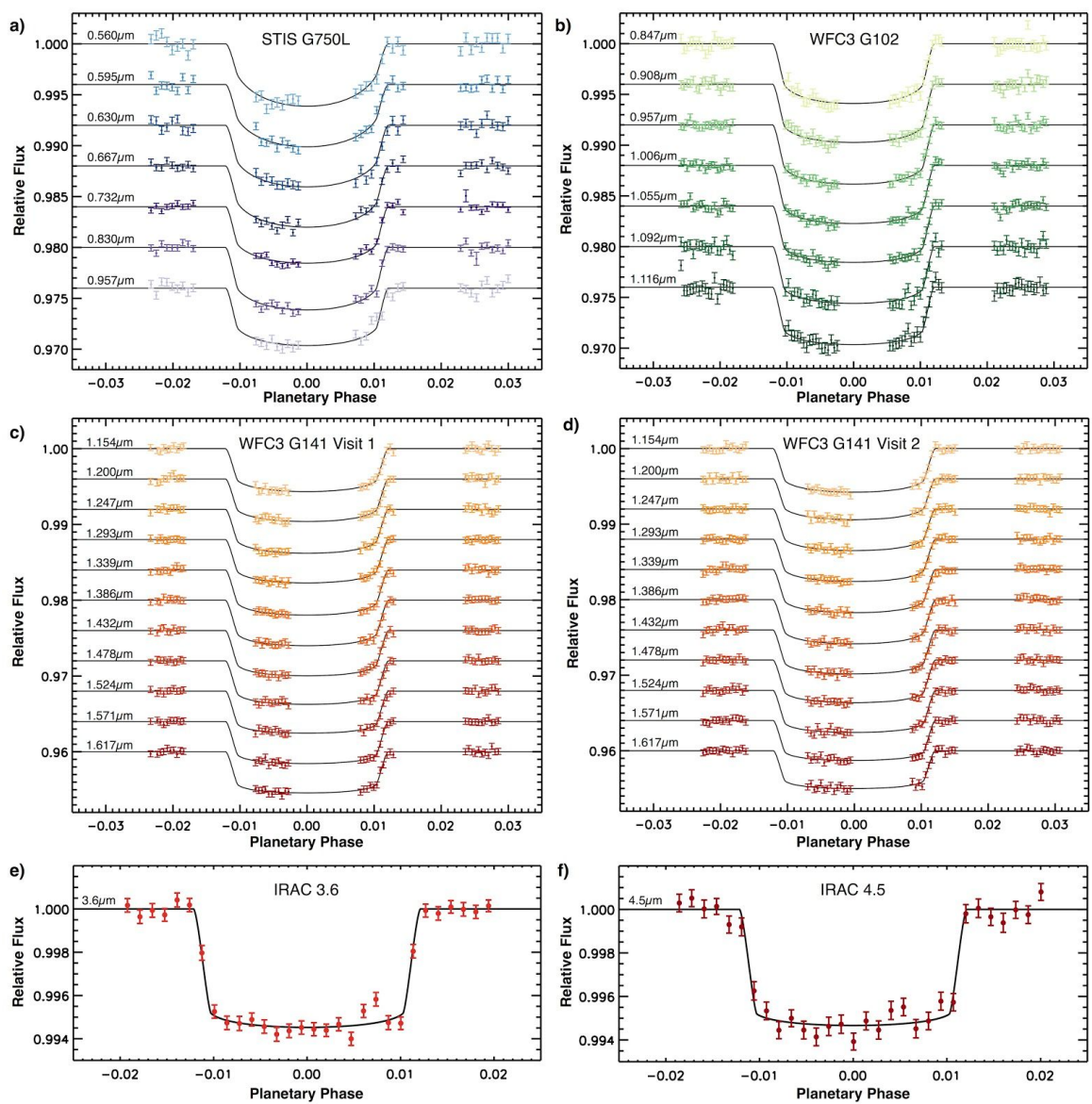

**Fig. S1. Spectroscopic light curves of all six transits of HAT-P-26b.** a) HST STIS, b) HST WFC3 G102 c) HST WFC3 G141 visit 1, d) G141 visit 2, e) Spitzer IRAC 3.6, and f) IRAC 4.5. Normalized and systematics-corrected data (colored points) are shown using the highest weighted systematic model for 7 spectroscopic channels spread in STIS from 0.5-1.0 μm, 7 channels for WFC3 G102 from 0.8-1.15 μm, 11 channels for WFC3 visit 1 & 2 from 1.1-1.7 μm, and Spitzer 3.6 & 4.5 μm photometric channels. Spitzer photometric light curves have been binned in time by ~2 minutes. Each light curve is offset and labeled for clarity.

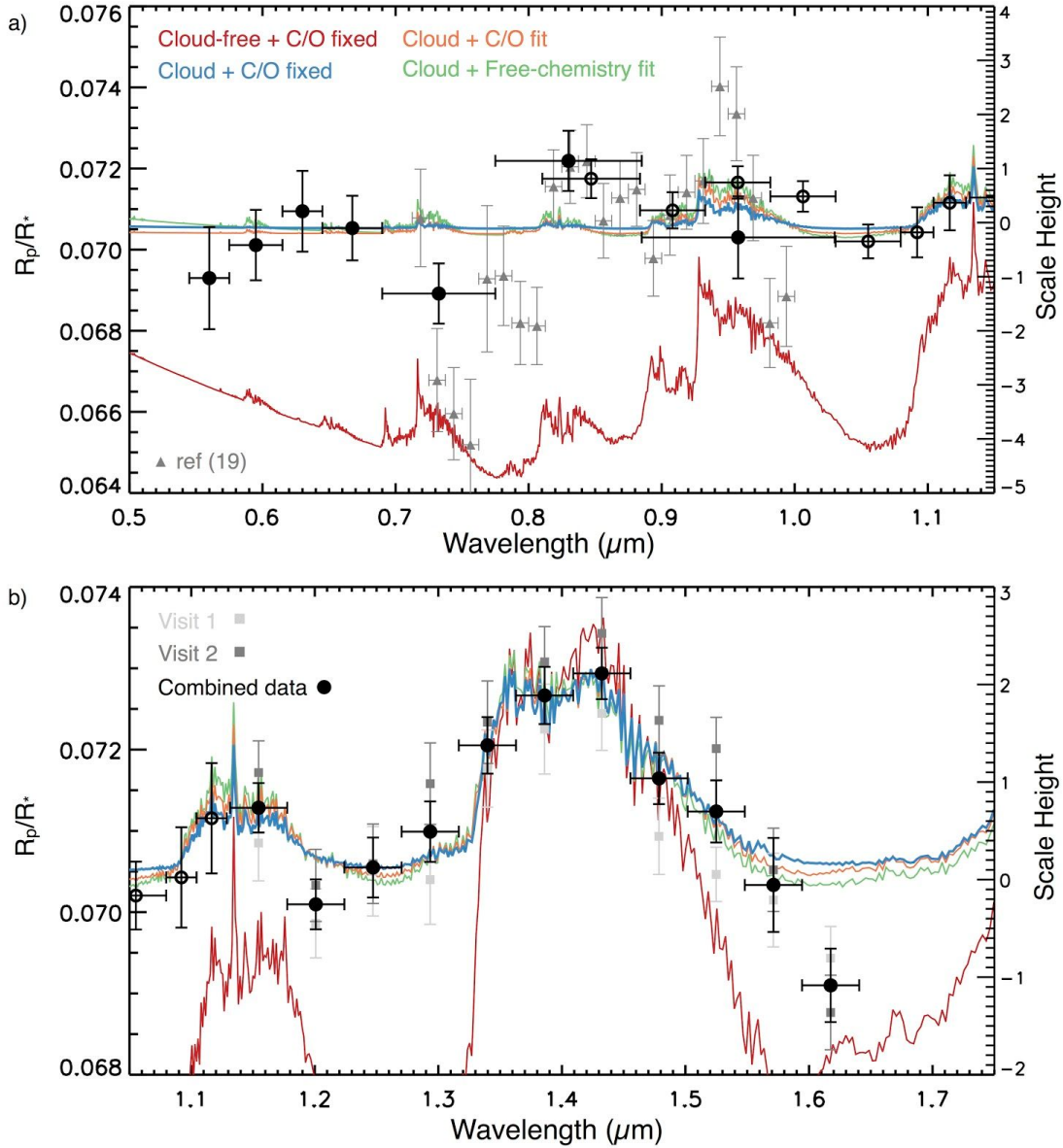

**Fig. S2. Expanded view of HAT-P-26b transmission spectrum from each HST visit.** Transmission spectrum measured for HAT-P-26b expanded to show the optical and near-infrared regions of the spectrum separately. Each plot shows models for four of the 12 ATMO retrievals (M1-3, M7; see Table S2 for the statistics). a) Optical STIS and WFC3 G102 measurements plotted with previous ground-based data (*19*). b) Prominent water absorption measured with WFC3 G141 over two visits shown as squares, with the final combined spectrum as solid black points. The right-hand axis shows the corresponding scale of the atmospheric transmission in terms of planetary scale height (as in Figure 1).

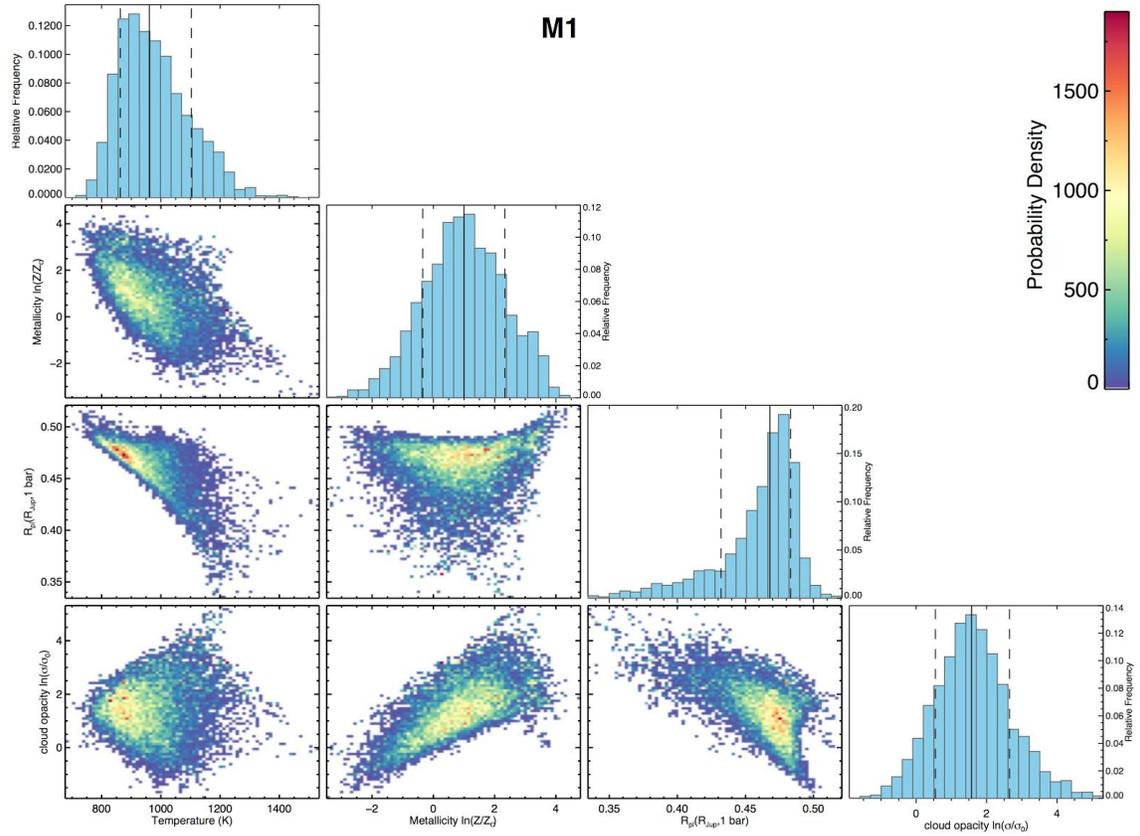

**Fig. S3**. **Correlation plot for pairs of retrieved parameters from the ATMO retrieval.** Correlation plot for model M1, which fits for the parameters, temperature, metallicity, radius ratio, and cloud opacity at a fixed solar C/O ratio and equilibrium chemistry. The posterior distributions show the probability density distribution of each fit. The vertical lines in the histograms indicate the fit value and uncertainty.

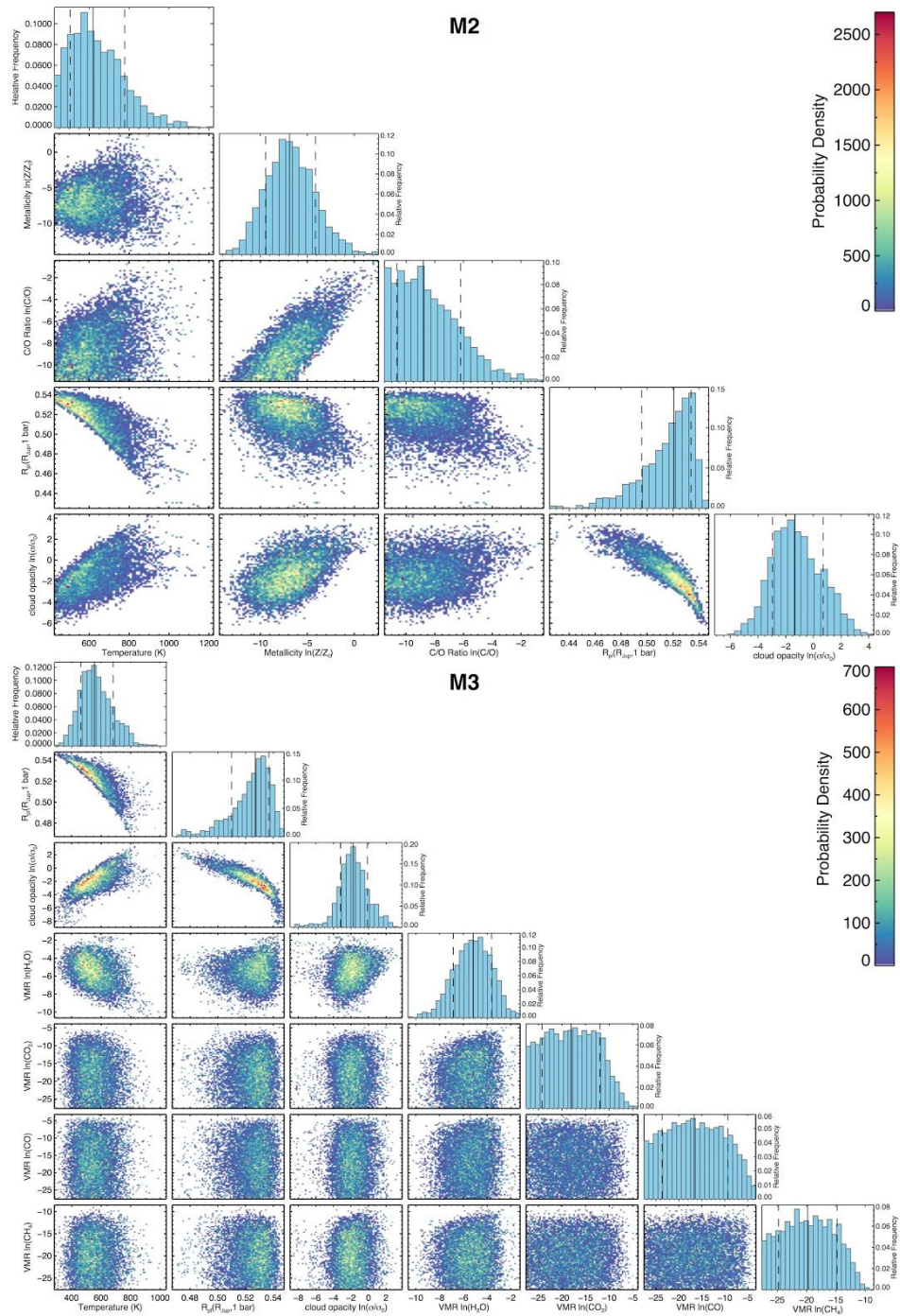

**Fig. S4. Correlation plot for pairs of retrieval parameters for models M2 and M3.** As in Fig S3 but for M2 and M3. M2 is the same as M1 but also fits C/O and M3 is a free-chemistry, isothermal model using the ATMO retrieval.

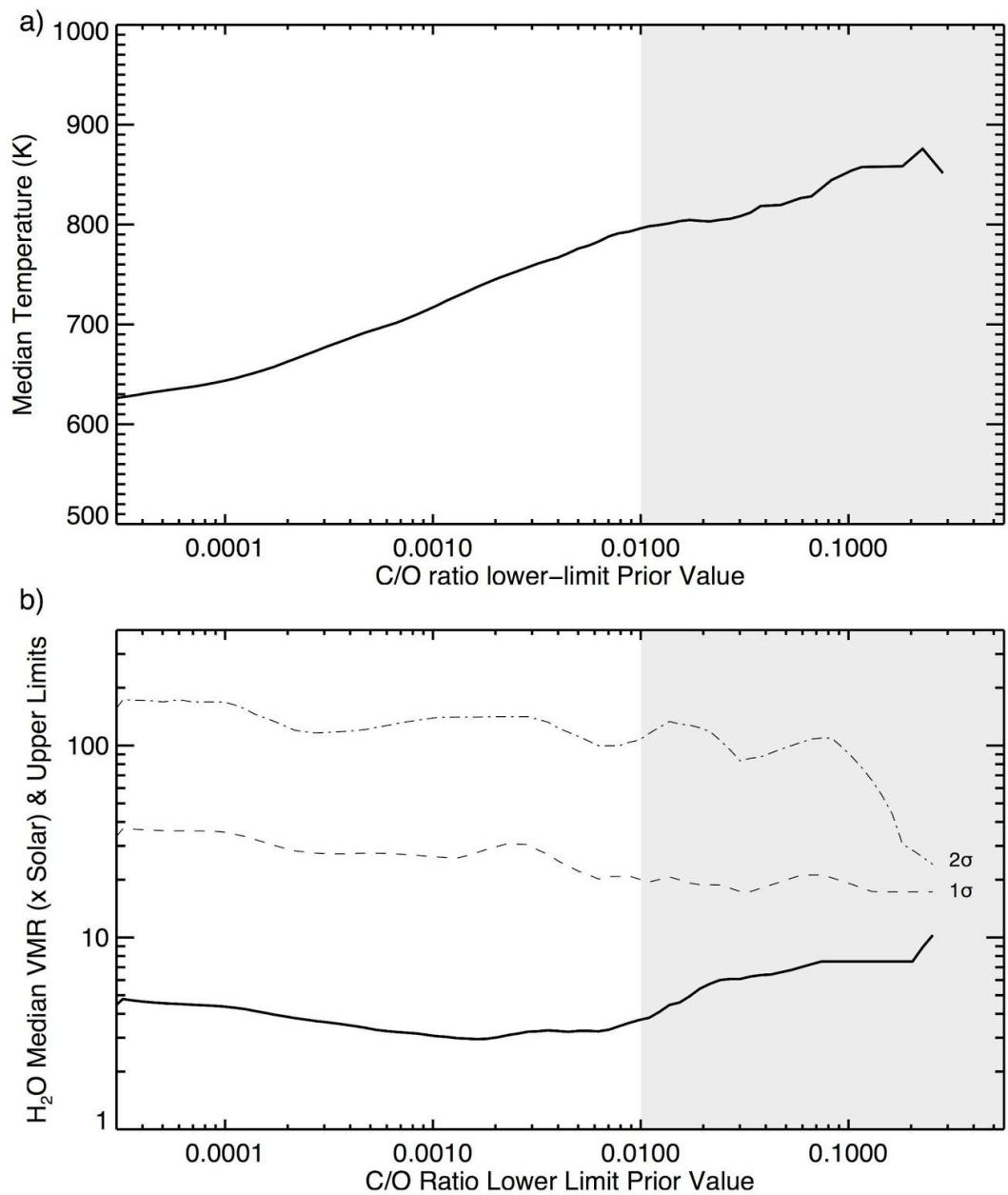

**Fig. S5**. **C/O prior impacts on the temperature and volume mixing ratio.** From model M2, a) the impact of a C/O prior on derived atmospheric temperature. b) impact of the C/O prior on the derived water abundance with the 1- and 2-σ upper limits. The grey shaded region in both plots shows the parameter space explored with the adopted prior constraint of C/O > 0.01.

**Table S1. HAT-P-26 system parameters fixed and derived.** The radius ($R^*$) and mass ($M^*$) of the star are in terms of solar radius and mass. Teff is the effective temperature of the star. We list the planetary radius ($R_p$) and mass ($M_p$) are in terms of Earth radius and mass planetary with gravity ($g_p$), and the derived inclination and semi-major axis in terms of stellar radius ($a/R^*$).

| Parameter | Value |
|---|---|
| Stellar Parameters | |
| $R_*$ ($R_{sun}$) | 0.788(-0.043 +0.098) ‡ |
| $M_*$ ($M_{sun}$) | 0.816 (±0.033) ‡ |
| $T_{eff}$ (K) | 5011(±55) ‡ |
| [Fe/H] | 0.01(±0.04) ‡ |
| Planet Parameters | |
| $R_p$ ($R_{Earth}$) | 6.33 † |
| $M_p$ ($M_{Earth}$) | 18.6 † |
| Period (days) | 4.2345 † |
| $g_p$ (ms$^{-2}$) | 4.47 † |
| eccentricity | 0.124 † |
| inclination (°) | 88.09 (± 0.553) |
| $a/R_*$ | 11.89 (± 0.417) |
| Derived Planetary Metallicity | |
| ln[*M/H*] | 1.566 (±1.7034) |
| Measured center of transit time (BJD$_{TBD}$) | |
| STIS G750L | 2457413.432836 (±0.000172) |
| WFC3 G102 | 2457616.690103 (±0.000011) |
| WFC3 G141 Visit 1 | 2457460.013266 (±0.000016) |
| WFC3 G141 Visit 2 | 2457510.827100 (±0.000016) |
| IRAC 3.6 | 2456545.361830 (±0.000030) |
| IRAC 4.5 | 2456405.623110 (±0.000070) |

† . system parameters adopted from (*1*)
‡ . Updated stellar parameters from (*52*)

**Table S2. Retrieval parameters and fits for all 12 model retrievals.** For each model we list the number of free parameters, BIC, and weight, which are used to calculate the marginalised metallicity of the atmosphere. Metallicity is listed as natural logs, where solar metallicity equals ln(1).

| Model | ln(Metallicity) | 1σ Uncertainty | # of free parameters | BIC | $W_q$ |
|---|---|---|---|---|---|
| M1 | 1.00 | 1.3391 | 4 | 58.32 | 0.425 |
| M2 | 1.72 | 1.7231 | 5 | 58.21 | 0.449 |
| M3 | 2.81 | 1.5995 | 7 | 62.13 | 0.063 |
| M4 | 2.14 | 1.0764 | 6 | 65.03 | 0.015 |
| M5 | 2.86 | 1.5952 | 7 | 65.43 | 0.012 |
| M6 | 2.84 | 1.5463 | 9 | 68.85 | 0.002 |
| M7 | 6.05 | 0.3080 | 3 | 77.09 | 0.000 |
| M8 | -0.30 | 1.2775 | 4 | 67.28 | 0.005 |
| M9 | 4.65 | 1.2242 | 6 | 64.05 | 0.024 |
| M10 | 6.08 | 0.3447 | 5 | 80.48 | 0.000 |
| M11 | -0.27 | 0.9195 | 6 | 69.19 | 0.002 |
| M12 | 3.80 | 1.9600 | 8 | 69.30 | 0.002 |
| **BMA** | **1.566 (4.8× solar)** | **1.7034** | | | |